\newlength\smallfigwidth
 \documentclass[showkeys,twocolumn,floatfix,showpacs,amsmath,amssymb,superscriptaddress,blindtext,a4paper]{revtex4-1}

\usepackage{cancel}
\usepackage[toc,page]{appendix}
\usepackage{graphicx}
\def\ba{\begin{eqnarray}}
\def\ea{\end{eqnarray}}
\def\be{\begin{equation}}
\def\ee{\end{equation}}

\begin{document}

\title{Geometrically induced reversion of Hall current in a topological insulator cavity}

\author{W.H. Campos}
\email{warlley.campos@ufv.br}
\author{W.A. Moura-Melo}
\email{winder@ufv.br}
\author{J.M. Fonseca}
\email{jakson.fonseca@ufv.br}
\affiliation{ Departamento de F\'isica, Universidade Federal de Vi\c cosa, 36570-900,
Vi\c cosa, Brazil}

\begin{abstract}
An electric charge near the surface of a topological insulator induces an image magnetic monopole. Here, we show that if the topological insulator surface has a 
negative curvature, namely in the case of a semispherical cavity, the induced Hall current 
reverses its rotation as the electric charge crosses the semisphere geometric focus. Such a 
reversion is shown to be equivalent of inverting the charge of the image magnetic monopole. We also discuss upon the case of a semicylindrical cavity, where Hall current reversion appears to be feasible of probing in realistic experiments.
\end{abstract}




\maketitle

\date{\today}

\section{Introduction and Motivation}

Condensed matter states described by Landau's theory are characterized by order parameters 
which are well-behaved, except at phase transitions, where they are related to symmetry 
breaking in the material structure. Instead, topological insulators (TI's) 
\cite{Hasan, Review, Ando} constitute a recently discovered state of matter that presents 
topological order. In such materials the bulk is gapped as in conventional insulators, 
but they support gapless metallic surface states as a manifestation of topological 
order. These metallic states are protected by time reversal symmetry (TRS) and their stability 
is robust against non-magnetic impurities or mechanical deformations on the surface. In addition, the carriers motion is shown to obey spin-momentum locking, with their spins lying on the 
surface and pointing perpendicularly to momentum everywhere \cite{Fan-Zhang}.

Such surface states may acquire a gap whenever TRS is broken, for instance, by means of a magnetic perturbation (applied field and/or film coating). As a consequence, a superficial Hall
conductivity,  $\sigma_{xy}=(n+\frac{1}{2})\frac{e^2}{h}$, takes place (we use CGS units; $n$ is an integer, $e$ is the electronic charge, $h$ is the Planck constant). In addition, low energy topological insulating phase properties may be described in terms of the so-called  topological magneto-electric effect (TMEE) \cite{Review, Ando, Qi2008}, which is a (topological) ground-state response function \cite{Essim}.
Such an effect may be readily encompassed in 
the usual electrodynamics, keeping the Maxwell equations unaltered in form, but changing 
the constitutive relations according to \cite{Qi2008,Qi}:
\begin{equation}
\begin{split}
\textbf{D}&=\textbf{E}+4\pi \textbf{P}-2\alpha\textit{P}_3\textbf{B}\\
\textbf{H}&=\textbf{B}-4\pi\textbf{M}+2\alpha\textit{P}_3\textbf{E},
\end{split}
\end{equation}
where $\alpha=e^2/\hbar c \approx 1/137$ is the fine structure constant and $P_3$ is the electric-magnetic polarization 
\cite{Qi2008,Qi}. In conventional insulators $P_3=0$, while it reads $P_3= {\rm sign}[\textbf{M}\cdot\hat{\textbf{n}}]/2= \pm 1/2$ for a TI. The 
direction of the surface magnetization, $\textbf{M}$, determines the sign of $P_3$: + (-) if it points out (in) to the TI
surface ($\hat{\textbf n}$ is a unit vector normal to the surface). Physically, we clearly realize how TMEE dictates the unique TI electromagnetic behavior: an 
electric (and/or magnetic) field induces both an electric polarization and a magnetization. Namely, 
an electric field crossing the TI surface induces on its surface a Hall current given by \cite{Review, Qi2008}:
\begin{equation} 
\label{eq:jhall}
\textbf{J}_{\rm Hall}= P_3\frac{e^2}{h}\hat{\textbf{n}}\times\textbf{E}\,.
\end{equation}
This current clearly depends on the surface geometry, and its associated magnetic field may have special features. For instance, in a conical TI this current yields a topological electric polarization that depends on the cone aperture angle,  say, under the same external conditions, wider ($\delta>30^o$) and narrower ($\delta<30^o$) cones appear to polarize in opposite directions \cite{jaksoncone}. On the other hand, a point-like electric charge located near a TI flat surface induces a magnetic field 
resembling that produced by a magnetic monopole \cite{Qi}. It should be stressed that such an image magnetic monopole is not an elementary neither an emergent excitation at all: it rather comes to be an artificial particle that describes the physical effects associated to the Hall sources taking place in the surface of a TI, as described by TMEE. Therefore, the monopoles discussed here look like images of Dirac magnetic monopoles \cite{Dirac-1931-monopoles}. In addition, they should not be confused with the timely emergent monopoles observed in spin ice systems, even though in both cases the monopole appears attached to a physical string of dipoles. Actually, in spin ices, the original degrees of freedom (magnetic dipoles) are shown to {\em fractionalize} into isolated magnetic poles attached into pairs by physical strings \cite{spinice3D, spinice2D1, spinice2D2}.
\begin{figure*}[ht]
\includegraphics[width=0.9\textwidth]{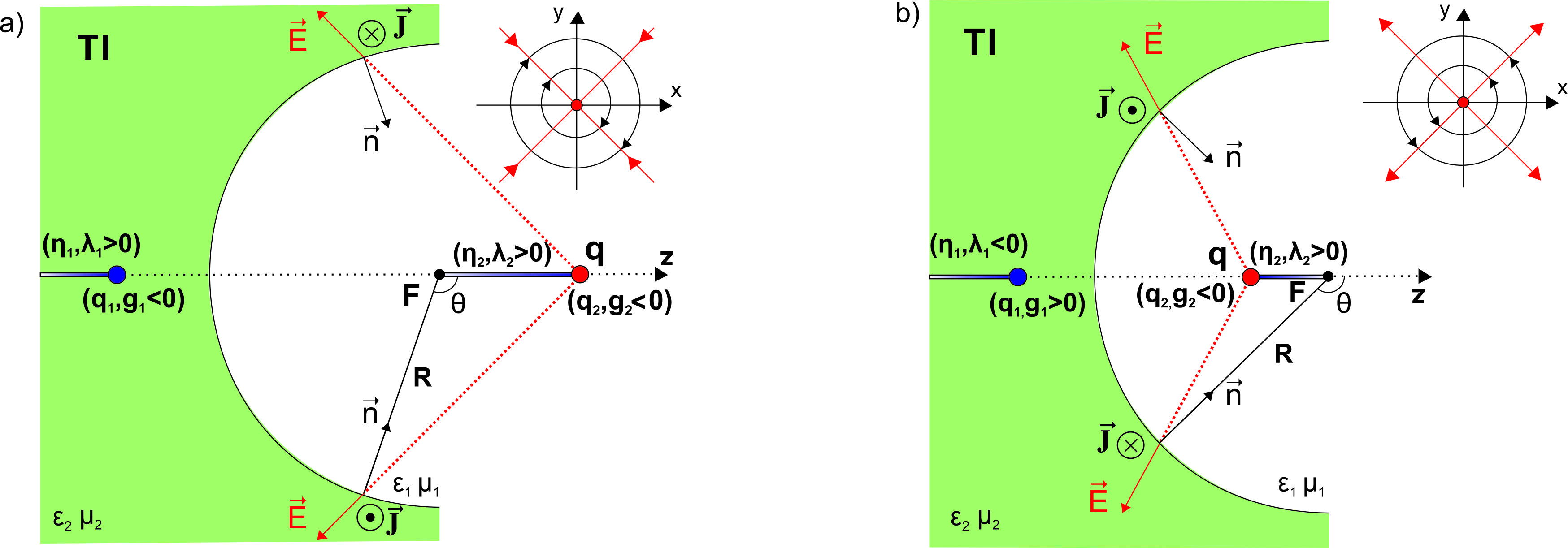}
\caption{[Color online] Schematic illustration of the semisphere cavity and its 
coordinate system. The z-axis origin is set at the focus \textbf{F},  $z_0$ accounts for the position of the 
electric charge, $q$, while ($\epsilon _1,\mu _1$) and ($\epsilon_2,\mu _2$)  are the electric permittivity 
and magnetic permeability of conventional and TI, respectively.
The magnetic monopole charge 
is reversed whenever the electric charge crosses the semisphere focus. In figure a), 
$z_0>0$, the Hall current, Eq. (2), rotates clockwise (see inset), 
being equivalent to a negative image monopole, $g_1<0$. In figure b), whenever $z_0<0$ the Hall current rotates 
counterclockwise, whose magnetic field resembles that produced by a monopole with charge $g_1>0$ (see inset). 
Electric charge $q$ was assumed to be positive, otherwise the Hall current and therefore all image charges experience a change of sign.
The red lines represent the electric field and the black circles represent the circulating surface current, 
$\textbf{J}_{\rm Hall}$. The magnetic field lines have been omitted for simplicity. Actually, the image object is a dyon, carrying both electric and magnetic charges ($q_1$ and $g_1$); attached to it there is a string of electric and magnetic charges, $\eta_1(z)$ and $\lambda_1(z)$ (see text for details). There is also a dyon ($q_2$,$g_2$) at $z_0$, attached to a string ($\eta_2(z)$,$\lambda_2(z)$); these (image) objects describes the magnetic field inside the TI. The latter dyon does not reverses its sign. $\theta$ is the usual spherical polar angle; due to the azimuthal symmetry of the cavity, its related angle, $\phi$, is not shown.}\label{emt2}
\end{figure*}
Here, we consider the case of a concave TI surface, namely, a semispherical cavity.[ The semicylindrical cavity is also discussed within some details]. Besides of realizing the induced image monopole, we also find that whenever the electric charge crosses the semispherical focus the magnetic charge of the monopole is reversed. Physically, such a picture corresponds to a reversion in the rotation of the induced Hall current whenever the electric charge exactly crosses the focus. Our paper is organized as 
follows: In Section 2 we present the problem and its geometry. It is analytically solved up to first order in $\alpha$. Section 3 is devoted to a better understanding of the image monopole configuration, once the monopole appears attached to a string carrying both electric and magnetic charges. Finally, we point out our Conclusions and Prospects for 
forthcoming works. [An Appendix is also provided in order to clarify some technical details].  
\section{The semispherical geometry and the monopole reversion} \label{sec:semispherical}
Let the left half-space, $(z<0)$, be a TI with an electric permittivity and magnetic permeability $(\epsilon _2,\mu _2)$ whereas the right side, $(z>0)$, is occupied by an ordinary insulator with $(\epsilon_1,\mu _1)$. A semispherical cavity, radius \textbf{R} centered at $z=0$, is made in the TI, according to Fig. \ref{emt2}. Let also the electric charge, $q$, be placed along $z$-axis, say, at $z_0$. Whenever one breaks TRS on the cavity surface, for instance, by coating it with a thin ferromagnetic film, a Hall current is induced by the electric field produced by $q$, according to Eq. (\ref{eq:jhall}), as depicted in Fig. \ref{emt2}. Note that the magnetic field produced by such a current is equivalent to that generated by an image magnetic monopole, $g_1$, along with an image magnetic string extending from it to $z\to -\infty$. However, as a whole the Hall current produces both, electric and magnetic fields, in such a way that the induced image carries both electric and magnetic charges, $(q_1,g_1)$: it resembles a {\em dyon}. In addition, there are also image strings carrying electric and magnetic charges extending from the dyon to $z\to-\infty$, accounted respectively by $\eta_1(z)$ and $\lambda_1(z)$ in Fig.\ref{emt2}. The physical reason why such strings must be in order comes from the fact that $\nabla \cdot\textbf{B}=0$, so that the total magnetic charge of the image string must exactly cancel the monopole charge: $\int \lambda_1(z)dz=-g_1$. In order to describe the magnetic field inside TI, there is also an image \emph{dyon} ($q_2$, $g_2$) located at $z_0$ and attached to a string ($\eta_2(z)$, $\lambda_2(z)$) extending from it to the semisphere focus {\bf F}. This picture is analogous to that realized in planar and spherical topological insulators, as reported in Ref. \cite{Qi}.

The main finding here concerns the reversion of the magnetic charge according to the location of the electric charge: if $q$ is placed at $z>0$ (on the right of the focus, \textbf{F}) $\textbf{J}_{\rm Hall}$ rotates clockwise, which corresponds to a negative magnetic monopole, $g_1<0$, panel a) in Fig.\ref{emt2}; whereas for $z_0<0$ the current flows counter-clockwise, whose magnetic field resembles that produced by a positive monopole, $g_1>0$, as depicted in panel b) Fig.\ref{emt2}. The electric charge of the dyon, $q_1$, does not experience any inversion in its sign as charge $q$ crosses the focus, neither do the image charges inside the cavity.

In order to show the results above in further details, let us consider the simple situation where both dielectrics have the same $(\epsilon, \mu)$, say, $\mu_1=\mu_2=\epsilon_1=\epsilon_2=1 $. In addition, consider only the first order contribution to the electric field, say, that due to the charge $q$: $\textbf{E}=q\frac{\vec{r}-z_0\hat{z}}{\arrowvert\vec{r}-z_0\hat{z}\arrowvert^3}$. Thus, Hall current is obtained, up to order $\alpha=e^2/\hbar c$ (linear response regime), to read:

\begin{equation} \label{eq:jhall2}
\textbf{J}_{\rm Hall}=-\frac{\alpha c}{4\pi}\frac{qz_0\sin\theta}{(R^2+{z_0}^2-2Rz_0\cos\theta)^{3/2}}\hat{\phi}\,,
\end{equation}

\noindent where we have used $P_3=+1/2$. Higher order contribution, ${\cal O}(\alpha^2)$, only correct the current by very small values and have been neglected.

\begin{figure}[!t]
    \includegraphics[width=0.49\textwidth]{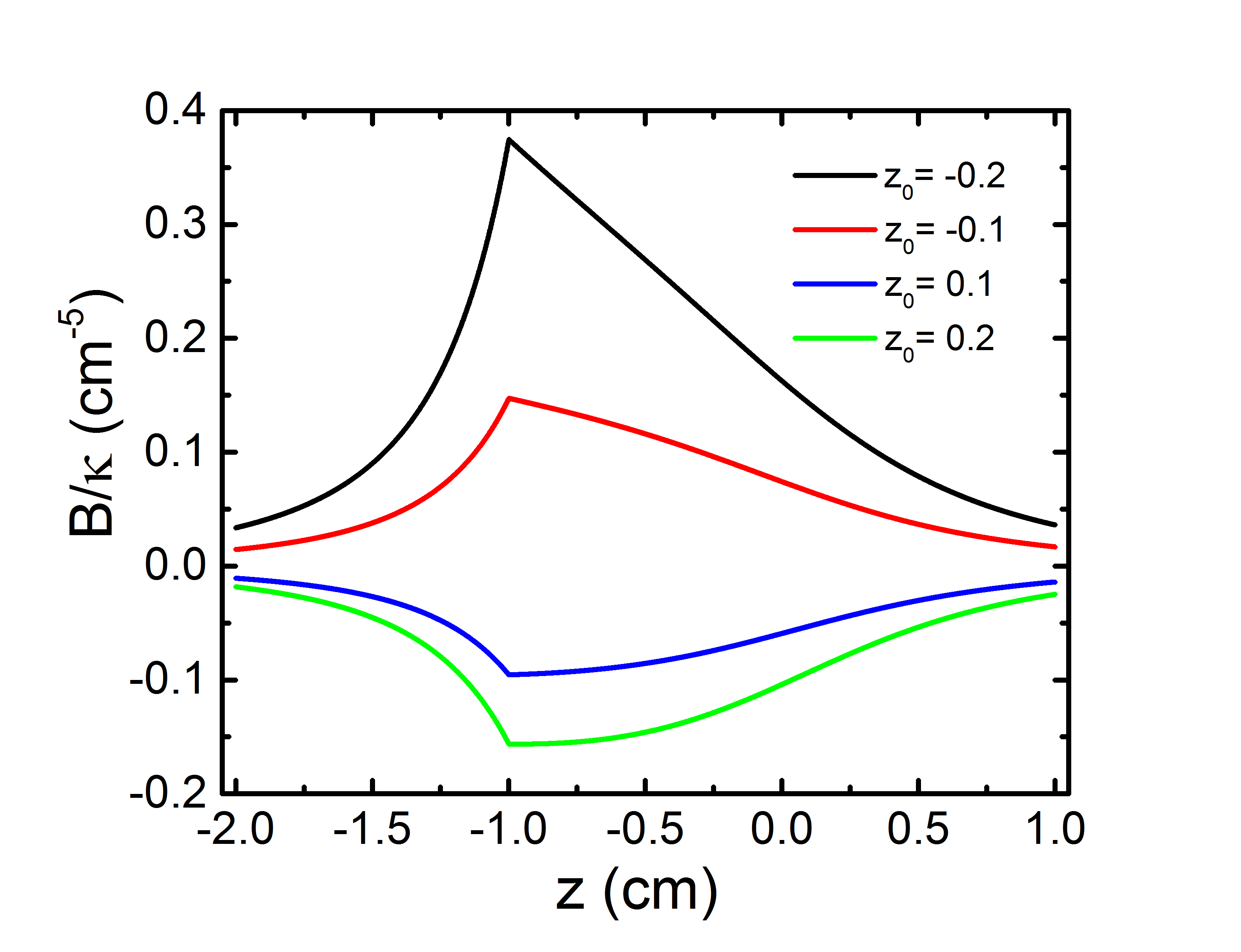}
    \caption{[Color online] Behavior of the magnetic field along $z$-axis, for $q$ fixed at different positions, $z_0$. Namely, note the change in the field sign whenever the electric charge crosses the cavity focus, $z=0$ (we have set $R=1 {\rm cm}$, and $\kappa=2\pi \dfrac{e^2}{2h}\dfrac{q}{c}R^3$).}
    \label{frev}
\end{figure}

\begin{figure}[!h]
	\includegraphics[width=0.23\textwidth]{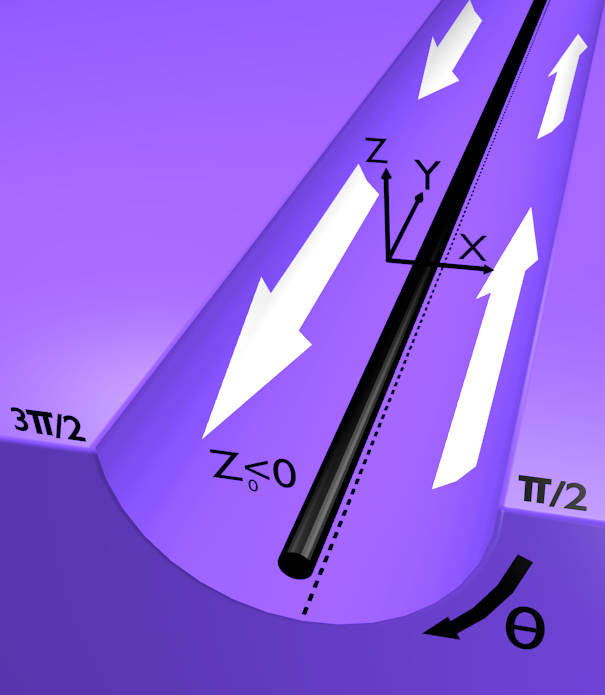}
	\includegraphics[width=0.23\textwidth]{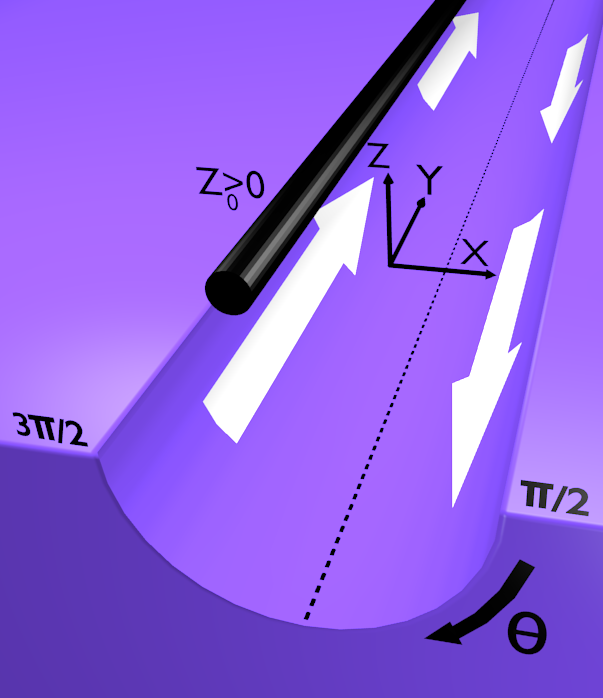}
	\caption{[Color online] The semicylindrical cavity, along with a straight wire of charge density (black bars along $y$-axis). In this case, the induced Hall current splits into two branches flowing in opposite directions, as indicated by white arrows. As the wire height crosses the cavity focus, $z_0=0$, both branches reverse their flows.}
	\label{semicylinder}
\end{figure}

From the expression above, one clearly realize the magnetic charge reversion whenever the charge $q$ crosses the cavity focus, $z_0=0$. Taking this current to the Biot-Savart equation, we obtain its generated magnetic field, ${\bf B}$, Eq. (\ref{magneticfield}). Such a calculus is tedious and length; it is presented in the Appendix. How ${\bf B}$ behaves along $z$-axis is shown in Fig. (\ref{frev}) for several 
values of $z_0$. Notice the change in the magnetic field when the electric charge is on the right or left of the focus, evidencing the reversion of the magnetic charge.
For the sake of completeness, let us also briefly discuss the case of a a semicylindrical cavity. Instead of a unique point charge, let also a wire carrying charge density $\lambda$ extending above the cavity along $y$-axis, as depicted in Fig.\ref{semicylinder}. The wire height $z_0$ may be varied and as it crosses $z_0=0$ each branch of the induced Hall current reverses its direction along the wire axis. Electric contacts at the cavity borders may directly measure the current properties and its reversion, as well. Therefore, while the straight wire splits the current into two longitudinal branches, the negative curvature of the cavity encompass its flow reversion, like occurs in the semisphere.
Actually, the induced Hall current is readily obtained to be (up to $\alpha^1$):
\begin{equation}\label{eq:jhall3}
\mathbf{J}_{\rm Hall}=-\frac{\alpha c}{4\pi}\frac{(2\lambda)z_0 \sin\theta}{R^2+z_0^2-2Rz_0\cos\theta}\,\hat{y}\,.
\end{equation}

\begin{figure}[!h]
	\includegraphics[width=0.5\textwidth]{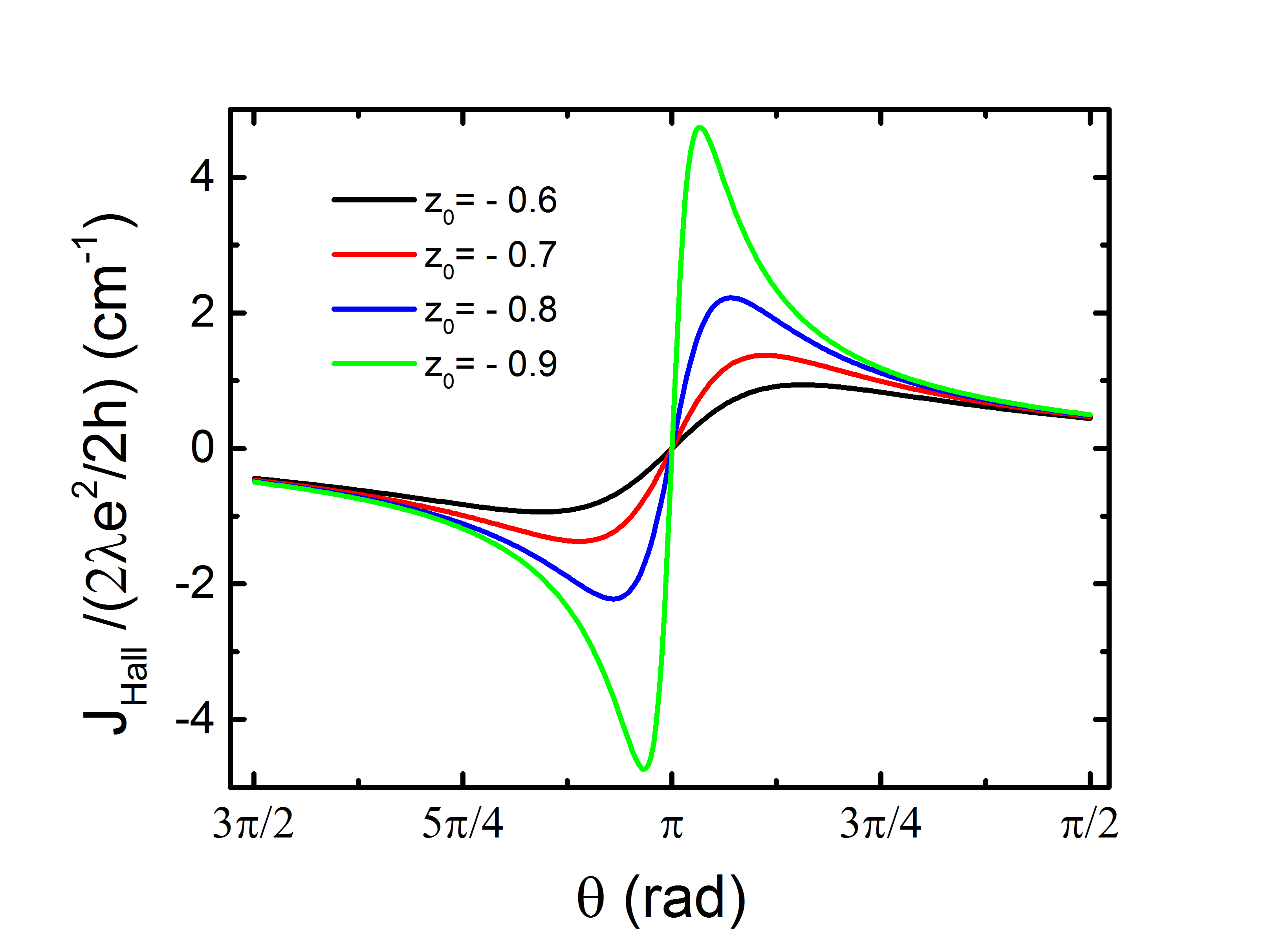}
	\caption{[Color online] How the Hall current behave. Left branch corresponds to $\pi <\theta\leq 3\pi /2$ while $\theta \in [\pi /2,\pi)$ accounts for the right branch (Fig. \ref{semicylinder}). We have taken a unity radius for the cavity, $R=1$, and put the wire in a number of heights inside the cavity ($z_0<0$).}
	\label{cylindercurrent}
\end{figure}

The above-mentioned current reversion may be easily realized from the expression above by setting $z_0\to-z_0$. The current behavior is plotted against $\theta$ in Fig. \ref{cylindercurrent}.
From a magnetic monopole point of view, this can be understood as a reversion of the monopole charge, analogously to what occurs in the semispherical case. Such a similarity may be clearly realized if we consider the 2D slice of both cases, giving us a semicircle problem.
\section{Magnetic monopole picture}\label{sec:magnetic monopole}
We shall discuss on the magnetic monopole description of the problem taking into account the semispherical cavity results. A similar picture may be done for semicylindrical cavity in an analogous way.

\begin{figure}[!h]
   \includegraphics[width=0.5\textwidth]{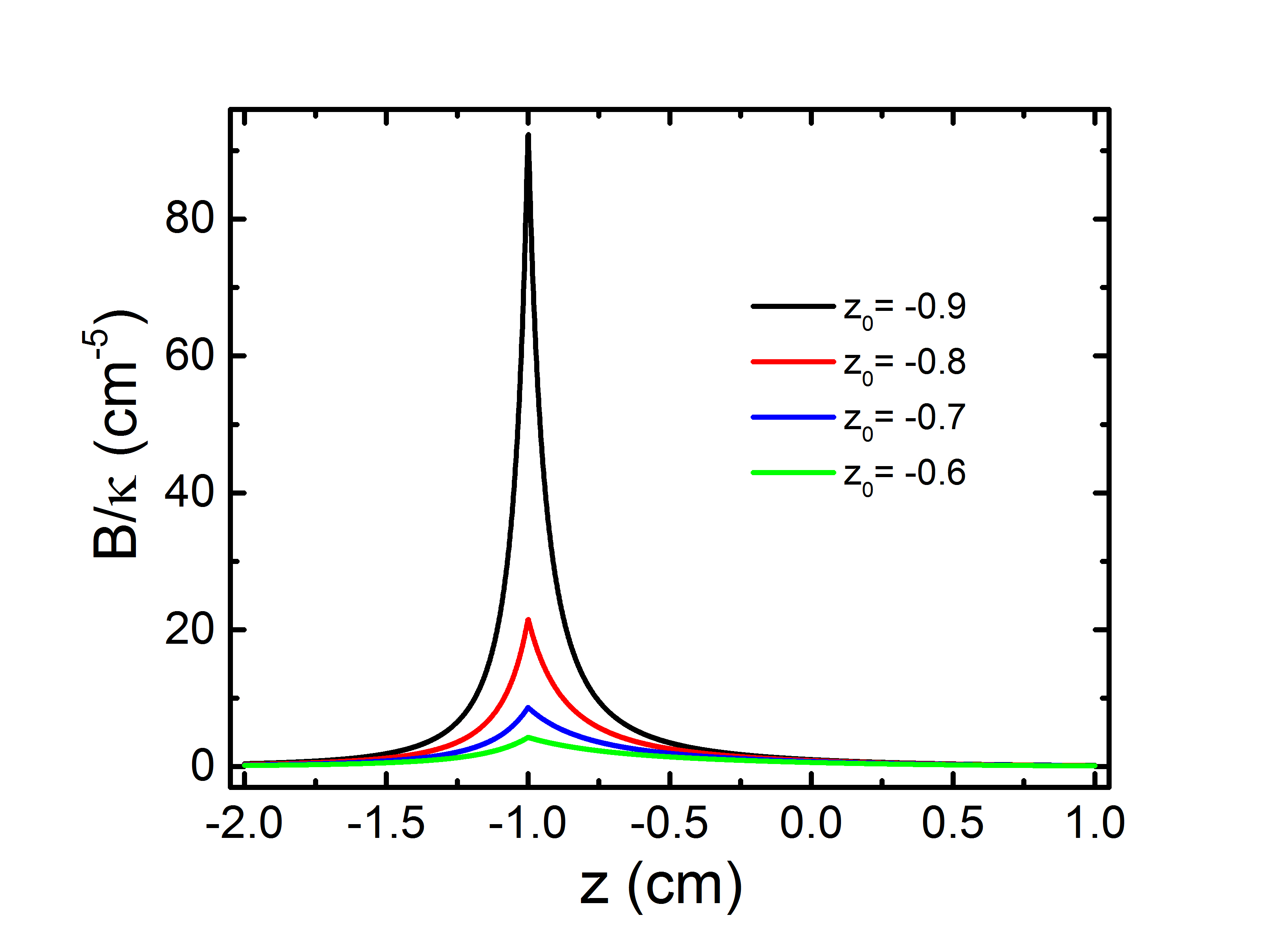}
    \caption{[Color online] Whenever the electric charge is placed very close to the interface, the experienced magnetic field is essentially that produced by an image point-like magnetic charge. Note the large increasing in the magnitude of ${\bf B}$ as one places the charge near to the interface, $z_0\sim -1$.}
    \label{plapp}
\end{figure}

The peaks at the interface, $z=-1$, presented in Fig. \ref{frev} are clear evidences of the monopole, and may be better understood whenever $q$ is taken close to $z=-1$, as shown in Fig. \ref{plapp}. Therefore, near the interface one experiences a magnetic field largely dominated by the monopole, so that as $z\to-1$ one practically realize only the point-like monopole itself. This is an expected result, since whenever one approaches the interface one should recover the result of a plane TI discussed in \cite{Qi}, say: charge $q$ induces an image point-like magnetic monopole. Essentially, the contribution due the image magnetic string is very small. Geometrically speaking, whenever $q$ is close to the interface, their relative separation is much smaller than the cavity curvature radius, say $(R-|z_0|)<<R$, then the surface seems to be flat.

\begin{figure}[!b]
  \includegraphics[width=0.45\textwidth]{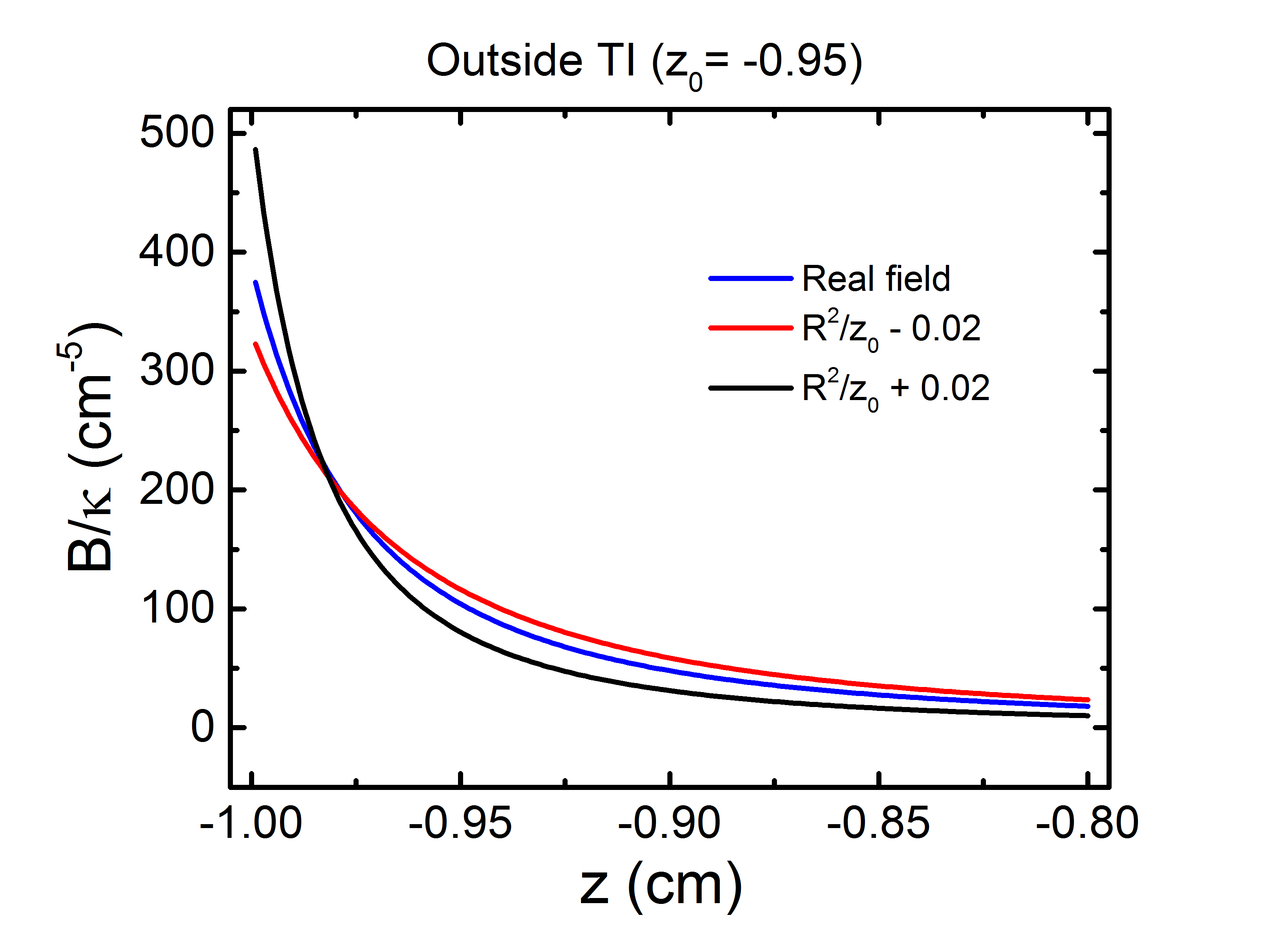}
    \caption{[Color online] Magnetic fields generated by the Hall current (blue curve), Eq. (\ref{magneticfield}), and by the image point-like magnetic monopole, $B_m=\mu_1 g_1/(z-z_m)^2$, slightly shifted from its exact position, $z_m$, by $\delta z_m=\pm 0.02 R$. We have taken $R=1\, {\rm cm}$ and placed $q$ close to the interface, at $z_0=-0.95 R$. As the magnetic monopole is placed at $z_m$ its field fits almost exactly the real curve.}
    \label{outti}
\end{figure}

In Fig. (\ref{outti}) we have plotted the magnetic field generated by the Hall current, Eq. (\ref{magneticfield}), with that produced by a unique point-like magnetic monopole, $B_m=\mu_1 g_1/(z-z_m)^2$. To improve the comparison, the monopole is slightly shifted around its exact position, $z_m=\frac{R^2}{z_0}$, by $\delta z_m=\pm 0.02 R$. At each of these shifted positions, the field due to the monopole alone approaches the real field, Eq. (\ref{magneticfield}). Actually, if the monopole is taken to $z_m$ its field fits that due to the induced Hall current with high precision. Such results further support the picture of the physical situation as that provided by a point-like magnetic monopole, namely, when we are close to the interface. But, how close does this picture remains valid? As wee shall see below, it is in good accordance even if the charge is taken to about $z_0 \sim -0.8 R$, provided that the probe is kept near to the interface, $z\approx -R$. 

\begin{figure}
    \includegraphics[width=0.5\textwidth]{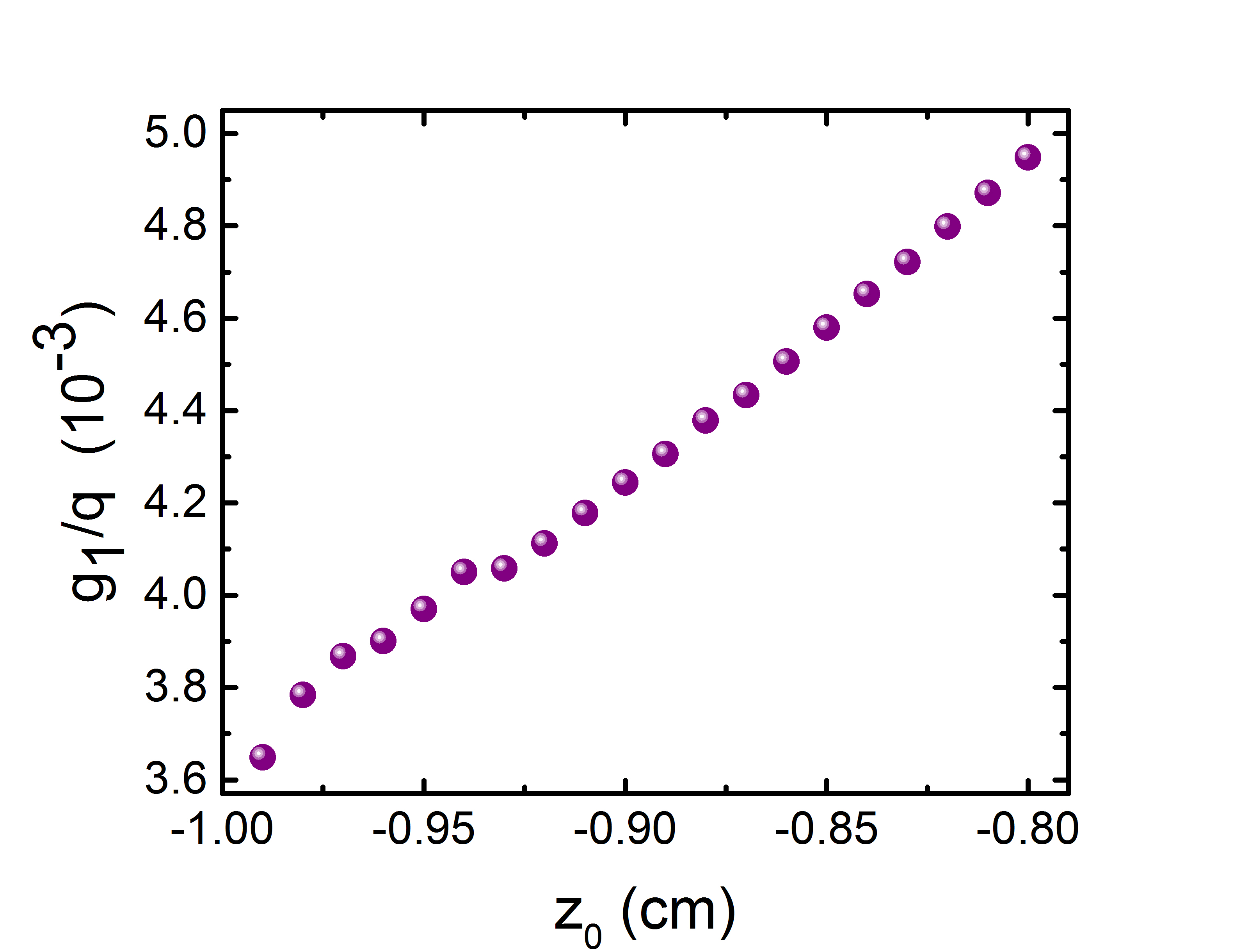}
    \caption{[Color online] Behavior of the image magnetic monopole strength as the electric charge is taken apart from the interface. Since monopole strength, $g_1$, linearly increases up to $z_0\approx -0.8 R$, the magnetic monopole picture is valid at least to such a separation, provided we also remain close to the interface (further details below).}
    \label{away}
\end{figure}

In Ref. \cite{Qi} the authors have found that the strength of the image magnetic monopole decreases with the inverse of the distance between the electric charge and 
the center of the spherical surface. Here, by virtue of the negative curvature of the semispherical cavity, the magnitude linearly increases, as shown in
Fig. (\ref{away}). Such a behavior occurs whenever $q$ is sufficiently close to the interface, say, the regime in which the point-like magnetic picture of the real problem is applicable. According to Fig. (\ref{away}), such a picture remains to be valid at least up to $z_0\sim -0.8R$. In particular, such an interval comprises the results presented in Fig. \ref{outti}.

However, if we get apart from the interface, say, observing the system from a far away point, $z>>R$, then 
magnetic field goes as $1/z^{3}$. This is not surprising, once as seen far away, the set magnetic string plus monopole looks like a magnetic dipole around $z_m$. At some extent, this is what happens with the tip of a magnetic needle, as reported in Ref.\cite{Armand}.

On the other hand, Hall current field spreads throughout the bulk of the topological insulator, as well. Therefore, from the point of view of an observer located inside the bulk, the magnetic field seems to be that produced by an image magnetic monopole, $g_2$, located at $z_0$, along with a string, $\lambda_2$. Then, as a whole, such an observer realize that all the electromagnetic field resembles that generated by a dyon placed at $z_0$ along with a string carrying both electric and magnetic charges, but extending from the monopole to the focus.

Following the discussion made by K\"onig {\em et al} \cite{Konig}, we emphasize that in realistic 3D TI's experiments the thickness of the sample takes only a few hundred atomic layers (around a few dozens of nanometers), constituting therefore a thin film. In such a case, an external field applied to a border side may cross the bulk reaching the opposite surface of the film. This is important because each surface contributes with a single Dirac cone yielding a net Hall conductivity of $\sigma_{i}=(-1)^{i+1}(n_i+\frac{1}{2})\frac{e^2}{h}$, with $i=1,2$ accounting for upper and bottom surface, respectively. Taking the linear response approach, the total magnetic field is simply the superposition of those generated by each Hall current. 
\section{Conclusions and Prospects}
In summary, we have seen that an electric charge placed near to a semispherical topological insulator cavity induces a superficial Hall current which reverses its rotation whenever the charge crosses the 
semisphere focus. Such a current is known to generate a magnetic field analogous that produced by an image magnetic monopole. Thus, current reversion is shown to be equivalent to invert the sign of the magnetic monopole.

Experimentally, to probe both  the induced Hall current along with its reversion (monopole reversion) a gap must be open in the surface states of the TI in a such a way that the Fermi level is kept within the gap. An efficient way to achieve this is by doping the as-grown TI with $Mn$ and $Fe$ dopants \cite{chen}. 
Later, let the TI cavity be coated with a soft ferromagnetic insulator, say, whose magnetization points normal to its surface. [Alternatively, $Cr$-doped $Bi_2(Se_x\,Te_{1-x})_3$ ferromagnetic TI's could be used. In this case, the magnetic order to break TRS is intrinsically provided by the compound itself, without necessity of further coating. See Ref. \cite{Zhang-Science-339-1582-2013} for details]. Now, whenever placing the electric charge near the cavity, its induced magnetic field does modify the film magnetization whose profile is dependent on the  monopole parameters, like strength and sign. Therefore, magnetic force microscopy (MFM) may be used to probe such a magnetization pattern. For the sake of completeness, if we consider a cavity of radius $R=1\,{\rm cm}$, then an electric charge around $q=10^{15}e$ induces a magnetic field about $100 \,{\rm Gauss}$, which is enough to modify the magnetization of the film in a detectable way.

Along these lines, the use of the semicylindrical cavity, as illustrated in Fig. \ref{semicylinder}, may be much more feasible for experiments. In this case, it is relatively easy to move the wire to vary its height to the cavity, leading to Hall current reversion. Such a current, along with its reversion may be directly measured by gate potential connected to the cavity borders. Even more important, such a set up may be useful to detect the half-integer Hall conductivity in direct transport experiments, as pointed out in the work of Ref. \cite{Konig}. Usually, the electric contacts attached to the surface are not able to capture only the local properties of the Hall current, due its curl character around the diode generating the bias potential. Here, the current flows straight along the wire axis diminishing considerably curl effects.\\

{\centerline {\bf Acknowledgements}}

The authors would like to thank J.A. Redinz for providing useful references concerning semispherical geometry, and J.B.S. Mendes for insightful discussions about MFM technique. Referee is also acknowledged for rising the interesting question of the semicylindrical cavity. They are also grateful to CAPES, CNPq and FAPEMIG (Brazilian agencies) for the financial support.\\
\section*{Appendix: Magnetic Field for semispheric geometry}
\label{sec:Appendix}

To obtain the magnetic field along z axis, one uses the current given by 
Eq. (\ref{eq:jhall2}) in the Biot-Savart law \cite{Jackson}: 

\begin{equation}
	\textbf{B}(z\textbf{\^ z})=\frac{1}{c}\int_{\mathcal{S}}\frac{\textbf{J}_{\rm Hall}
	(\textbf{r}\mathrm{'})\times (z\textbf{\^ z}-\textbf{r}\mathrm{'})}{|z\textbf{\^ 
	z}-\textbf{r}\mathrm{'}|^{3}}da\mathrm{'},
\end{equation}
where $\mathcal{S}$ represents the semispherical interface, $da\mathrm{'}$ is the infinitesimal area 
element and $c$ is the speed of light. After some algebra, we obtain: 
\begin{eqnarray}
\textbf{B}(z\textbf{\^ z})= &-&
2\pi \frac{e^2}{2h}\frac{qz_0}{c}R^3\bigg\{ \int_{\frac{\pi}{2}}^{\pi}\frac{\sin\theta'}
{(a+b\cos\theta'+d\cos^2\theta')^{\frac{3}{2}}}d\theta' \nonumber\\ 
	&-& \int_{\frac{\pi}{2}}^{\pi}\frac{\sin\theta'\cos^2\theta'}{(a+b\cos\theta'+d\cos^2\theta')^{\frac{3}
	{2}}}d\theta' \bigg\}\textbf{\^ z}\,,
\end{eqnarray}
\noindent
where $a=(R^2+z_0^2)(R^2+z^2),\quad
b=-2R(zR^2+zz_0^2+z_0R^2+z_0z^2),\quad $and $
d=4R^2zz_0$.

The first integral may be readily solved, while the second one needs the result from Eq. (2.264-7), Ref. \cite{Grad}. The final expression takes the form below, which has been used in Sections 2 and 3, mainly for numeric evaluation of a number of plots. It should be stressed that such an expression is exact up to 1st order in $\alpha$ (linear response regime).

\onecolumngrid
\begin{equation}
\label{magneticfield}
\textbf{B}(z\textbf{\^ z}) = -2\pi \frac{e^2}{2h}\frac{qz_0}{c}R^3\textbf{\^ z}
	\begin{cases}
		\begin{split} \frac{4d-2b}{(4ad-b^2)\sqrt{a-b+d}}+&\frac{2b}{(4ad-b^2)\sqrt{a}}+\\+\frac{1}{d(4ad-b^2)}\bigg[&\frac{4ad-2b^2+2ab}{\sqrt{a-b+d}}-\frac{2ab}{\sqrt{a}}\bigg]+\\&+\frac{1}{d\sqrt{d}}\ln{\frac{2\sqrt{d(a-b+d)}-2d+b}{2\sqrt{ad}+b}} \end{split} & \mbox{if } d>0\ \\ \\
		\begin{split} \frac{4d-2b}{(4ad-b^2)\sqrt{a-b+d}}+&\frac{2b}{(4ad-b^2)\sqrt{a}}+\\+\frac{1}{d(4ad-b^2)}\bigg[&\frac{4ad-2b^2+2ab}{\sqrt{a-b+d}}-\frac{2ab}{\sqrt{a}}\bigg]+\\+\frac{1}{d\sqrt{-d}}\bigg[&\arcsin{\frac{b}{\sqrt{b^2-4ad}}-\arcsin{\frac{b-2d}{\sqrt{b^2-4ad}}}}\bigg]\end{split} & \mbox{if } d<0 \end{cases}\,.
\end{equation}
\twocolumngrid

\end{document}